\pdfoutput=1
\documentclass{entcs} \usepackage{entcsmacro}
\usepackage{graphicx}
\usepackage{bussproofs} %
\usepackage{tikz}

\newcommand{\Nat}{{\mathbb N}}

\def\lastname{Ramos, de Queiroz, de Oliveira}
\begin{document}
\begin{frontmatter}
  \title{On the Identity Type as the Type of Computational Paths} \author{Arthur F. Ramos\thanksref{myemail}}
  \address{Centro de Informática\\ Universidade Federal de Pernambuco\\
    Recife, Brazil} 
\author{Ruy J. G. B. de Queiroz\thanksref{coemail}}
  \address{Centro de Informática\\Universidade Federal de Pernambuco\\
  Recife, Brazil} 
\author{Anjolina G. de Oliveira\thanksref{cocoemail}}
  \address{Centro de Informática\\Universidade Federal de Pernambuco\\
  Recife, Brazil}
 \ \thanks[myemail]{Email:
    \href{afr@cin.ufpe.br} {\texttt{\normalshape
        afr@cin.ufpe.br}}} \thanks[coemail]{Email:
    \href{ruy@cin.ufpe.br} {\texttt{\normalshape
        ruy@cin.ufpe.br}}}
    \thanks[cocoemail]{Email:\href{ago@cin.ufpe.br} {\texttt{\normalshape
        ago@cin.ufpe.br}}}

\begin{abstract} 
We introduce a new way of formalizing the intensional identity type based on the fact that a entity known as computational paths can be interpreted as terms of the identity type. Our approach enjoys the fact that our elimination rule is easy to understand and use. We make this point clear constructing terms of some relevant types using our proposed elimination rule. We also show that the identity type, as defined by our approach, induces a groupoid structure. This result is on par with the fact that the traditional identity type induces a groupoid, as exposed by Hofmann \& Streicher (1994).
\end{abstract}
\end{frontmatter}
\section{Introduction}\label{intro}

One interesting peculiarity of  Martin-L\"of's Intensional Type Theory is the existence of two distinct kind of equalities between terms of the same type. The first one is originated by the fact that equality can be seen as a type. The second one is originated by the fact that two terms can be equal by definition.

The treatment of an equality as a type gives rise to a extremely interesting type known as identity type. The idea is that, given terms $a, b$ of a type $A$, one may form the type whose elements are proofs that $a$ and $b$ are equal elements of type $A$. This type is represented by $Id_{A}(a,b)$. A term $p : Id_{A}(a,b)$ makes up for the {\em grounds\/} \cite{Prawitz2009} (or proof) that estabilishes that $a$ is indeed equal to $b$. We say that $a$ is {\em propositionally\/} equal to $b$.

 The second kind of equality is called definitional and is denotated by $\equiv$. It occurs when two terms are equal by definition, i.e., there is no need for an evidence or a proof to estabilish the equality. A classic example, given in \cite{hott}, is to consider any function definition, for example, $f: \Nat \rightarrow \Nat$ defined by $f(x) \equiv x^{2}$. For $x=3$, we have, by definition, that $f(x) \equiv 3^{2}$. We are unable to conclude that $f(3) \equiv 9$ though. The problem is that, to conclude that $3^2 = 9$, we need additional evidence. We need a way to compute $3^2$, i.e., we need to use exponentials (or multiplications) to conclude that $3^2$ is equal to $9$. That way, we can only prove that $3^2$ is propositionally equal to $9$, i.e., there is a $p : Id_{\Nat}(3^{2},9)$. 

Between those two kind of equalities, the propositional one is, without a doubt, the most interesting one. This claim is based on the fact that many interesting results have been achieved using the identity type. One of these was the discovery of the Univalent Models in 2005 by Vladimir Voevodsky \cite{Vlad1}. A groundbreaking result has arisen from Voevdsky's work: the connection between type theory and homotopy theory. The intuitive connection is simple: a term $a : A$ can be considered as a point of the space $A$ and $p: Id_{A}(a,b)$ is a homotopical path between points $a, b \in A$ \cite{hott}. This has given rise to a whole new area of research, known as Homotopy Type Theory. It leads to a new perspective on the study of equality, as expressed by Voevodsky in a recent talk in {\em The Paul Bernays Lectures\/} (Sept 2014, Zurich): equality (for abstract sets) should be looked at as a {\em structure\/} rather than as a {\em relation\/}. 

Motivated by the fact that the identity type has given rise to such interesting concepts, we have been engaged in a process of revisiting the construction of the intensional identity type, as originally proposed by Martin-L\"of. Although beautifully defined, we have noticed that proofs that uses the identity type can be sometimes a little too complex. The elimination rule of the intensional identity type encapsulates lots of information, sometimes making too troublesome the process of finding the reason that builds the correct type.

Inspired by the path-based approach of the homotopic interpretation, we believe that a similar approach can be used to define the identity type in type theory. To achieve that, we have been using a notion of {\em computational paths}. The interpretation will be similar to the homotopic one: a term $p : Id_{A}(a,b)$ will be a computational path betweem terms $a, b : A$, and such path will be the result of a sequence of rewrites. In the sequel, we shall define formally the concept of a computational path. The main idea, i.e.\ proofs of equality statements as (reversible) sequences of rewrites, is not new, as it goes back to a paper entitled ``Equality in labelled deductive systems and the functional interpretation of propositional equality", presented in December 1993 at the {\em 9th Amsterdam Colloquium\/}, and published in the proceedings in 1994 \cite{Ruy4}.

\section{Computational Paths} \label{path}

Since computational path is a generic term, it is important to emphasize the fact that we are using the term computational path in the sense defined by \cite{Ruy5}. A computational path is based on the idea that it is possible to formally define when two computational objects $a,b : A$ are equal. These two objects are equal if one can reach $b$ from $a$ applying a sequence of axioms or rules. This sequence of operations forms a path. Since it is between two computational objects, it is said that this path is a computational one. Also, an application of a axiom or a rule transforms (or rewrite) an term in another. For that reason, a computational path is also known as a sequence of rewrites. Nevertheless, before we define formally a computational path, we can take a look at one famous equality theory, the $\lambda\beta\eta-equality$ \cite{lambda}:

\begin{definition}
The \emph{$\lambda\beta\eta$-equality} is composed by the following axioms:

\begin{enumerate}
\item[$(\alpha)$] $\lambda x.M = \lambda y.[y/x]M$ \quad if $y \notin FV(M)$; 
\item[$(\beta)$] $(\lambda x.M)N = [N/x]M$;
\item[$(\rho)$] $M = M$;
\item[$(\eta)$] $(\lambda x.Mx) = M$ \quad $(x \notin FV(M))$.
\end{enumerate}

And the following rules of inference:

\bigskip
\noindent
\begin{bprooftree}
\AxiomC{$M = M'$ }
\LeftLabel{$(\mu)$ \quad}
\UnaryInfC{$NM = NM'$}
\end{bprooftree}
\begin{bprooftree}
\AxiomC{$M = N$}
\AxiomC{$N = P$}
\LeftLabel{$(\tau)$}
\BinaryInfC{$M = P$}
\end{bprooftree}

\bigskip
\noindent
\begin{bprooftree}
\AxiomC{$M = M'$ }
\LeftLabel{$(\nu)$ \quad}
\UnaryInfC{$MN = M'N$}
\end{bprooftree}
\begin{bprooftree}
\AxiomC{$M = N$}
\LeftLabel{$(\sigma)$}
\UnaryInfC{$N = M$}
\end{bprooftree}

\bigskip
\noindent
\begin{bprooftree}
\AxiomC{$M = M'$ }
\LeftLabel{$(\xi)$ \quad}
\UnaryInfC{$\lambda x.M= \lambda x.M'$}
\end{bprooftree}




\end{definition}





\begin{definition}(\cite{lambda})
$P$ is $\beta$-equal or $\beta$-convertible to $Q$  (notation $P=_\beta Q$) 
iff $Q$ is obtained from $P$ by a finite (perhaps empty)  series of $\beta$-contractions
and reversed $\beta$-contractions  and changes of bound variables.  That is,
$P=_\beta Q$ iff \textbf{there exist} $P_0, \ldots, P_n$ ($n\geq 0$)  such that
$P_0\equiv P$,  $P_n\equiv Q$,
$(\forall i\leq n-1) (P_i\triangleright_{1\beta}P_{i+1}  \mbox{ or }P_{i+1}\triangleright_{1\beta}P_i  \mbox{ or } P_i\equiv_\alpha P_{i+1}).$
\end{definition}
\noindent (NB: equality with an \textbf{existential} force, which will show in the proof rules for the identity type.)

The same happens with $\lambda\beta\eta$-equality:\\ 
\begin{definition}($\lambda\beta\eta$-equality \cite{lambda})
The equality-relation determined by the theory $\lambda\beta\eta$ is called
$=_{\beta\eta}$; that is, we define
$$M=_{\beta\eta}N\quad\Leftrightarrow\quad\lambda\beta\eta\vdash M=N.$$
\end{definition}

\begin{example}
Take the term $M\equiv(\lambda x.(\lambda y.yx)(\lambda w.zw))v$. Then, it is $\beta\eta$-equal to $N\equiv zv$ because of the sequence:\\
$(\lambda x.(\lambda y.yx)(\lambda w.zw))v, \quad  (\lambda x.(\lambda y.yx)z)v, \quad   (\lambda y.yv)z , \quad zv$\\
which starts from $M$ and ends with $N$, and each member of the sequence is obtained via 1-step $\beta$- or $\eta$-contraction of a previous term in the sequence. To take this sequence into a {\em path\/}, one has to apply transitivity twice, as we do in the example \ref{examplepath} below.
\end{example}


The aforementioned theory estabilishes the equality between two $\lambda$-terms. Since we are working with computational objects as terms of a type, we need to translate the $\lambda\beta\eta$-equality to a suitable equality theory based on Martin L\"of's type theory. We obtain:
	
\begin{definition}
The equality theory of Martin L\"of's type theory has the following basic proof rules for the $\Pi$-type:

\bigskip

\noindent
\begin{bprooftree}
\hskip -0.3pt
\alwaysNoLine
\AxiomC{$N : A$}
\AxiomC{$[x : A]$}
\UnaryInfC{$M : B$}
\alwaysSingleLine
\LeftLabel{$(\beta$) \quad}
\BinaryInfC{$(\lambda x.M)N = M[N/x] : B$}
\end{bprooftree}
\begin{bprooftree}
\hskip 11pt
\alwaysNoLine
\AxiomC{$[x : A]$}
\UnaryInfC{$M = M' : B$}
\alwaysSingleLine
\LeftLabel{$(\xi)$ \quad}
\UnaryInfC{$\lambda x.M = \lambda x.M' : (\Pi x : A)B$}
\end{bprooftree}

\bigskip

\noindent
\begin{bprooftree}
\hskip -0.5pt
\AxiomC{$M : A$}
\LeftLabel{$(\rho)$ \quad}
\UnaryInfC{$M = M : A$}
\end{bprooftree}
\begin{bprooftree}
\hskip 72pt
\AxiomC{$M = M' : A$}
\AxiomC{$N : (\Pi x : A)B$}
\LeftLabel{$(\mu)$ \quad}
\BinaryInfC{$NM = NM' : B$}
\end{bprooftree}

\bigskip

\noindent
\begin{bprooftree}
\hskip -0.5pt
\AxiomC{$M = N : A$}
\LeftLabel{$(\sigma) \quad$}
\UnaryInfC{$N = M : A$}
\end{bprooftree}
\begin{bprooftree}
\hskip 76pt
\AxiomC{$N : A$}
\AxiomC{$M = M' : (\Pi x : A)B$}
\LeftLabel{$(\upsilon)$ \quad}
\BinaryInfC{$MN = M'N : B$}
\end{bprooftree}

\bigskip

\noindent
\begin{bprooftree}
\hskip -0.5pt
\AxiomC{$M = N : A$}
\AxiomC{$N = P : A$}
\LeftLabel{$(\tau)$ \quad}
\BinaryInfC{$M = P : A$}
\end{bprooftree}

\bigskip

\noindent
\begin{bprooftree}
\hskip -0.5pt
\AxiomC{$M: (\Pi x : A)B$}
\LeftLabel{$(\eta)$ \quad}
\RightLabel {$(x \notin FV(M))$}
\UnaryInfC{$(\lambda x.Mx) = M: (\Pi x : A)B$}
\end{bprooftree}

\bigskip

\end{definition}

We are finally able to formally define computational paths:

\begin{definition}
Let $a$ and $b$ be elements of a type $A$. Then, a \emph{computational path} $s$ from $a$ to $b$ is a sequence of rewrites (each rewrite is an application of the inference rules of the equality theory of type theory or is a change of bound variables). We denote that by $a =_{s} b$.
\end{definition}

The definition makes clear the power of the computational paths. Since each rewrite is an application of intuitive axioms, working with computational paths is straightfoward. In addition, the presence of axioms such as the reflexivity ($\rho$), the transitivity ($\tau$) and the symmetry($\sigma$) opens the possibility of computational paths inducing a groupoidal structure. This will be the case, as we will see further in this work.

\section{Identity Type}

As we have already mentioned, our objective is to propose a formalization to the identity type using computational paths. We also want to make clear that we are working with the intensional version of the identity type. We assure that due to the fact that there exists an extensional version. Nevertheless, this extensional version does not have  most of the interesting properties that the intensional one has (the homotopic interpretation, for example). Since our approach is based on computational paths, we will sometimes refer to our formulation as the \emph{path-based} approach and the traditional formulation as the \emph{pathless} approach.  Before the deductions that build the path-based identity type, we would like to show the construction of the traditional approach \cite{harper1}:

\bigskip
\begin{bprooftree}
\AxiomC{$A$ type}
\AxiomC{$M : A$}
\AxiomC{$N : A$}
\RightLabel{$Id- F$ \quad}
\TrinaryInfC{$Id_{A}(M,N)$ type}
\end{bprooftree}
\begin{bprooftree}
\AxiomC{$a : A$}
\RightLabel{$Id - I$ \quad}
\UnaryInfC{$r(a) : Id_{A}(a,a)$}
\end{bprooftree}
\bigskip
\begin{center}
\begin{bprooftree}
\alwaysNoLine
\AxiomC{$M:A$}
\AxiomC{$N:A$}
\AxiomC{$P:Id_{A}(M,N)$}
\AxiomC{$[x:A]$}
\UnaryInfC{$Q(x):C(x,x,r(x))$}
\AxiomC{$[x:A,y:A,z:Id_{A}(x,y)]$}
\UnaryInfC{$C(x,y,z)$ type} 
\RightLabel{$Id - E$ \quad}
\alwaysSingleLine
\QuinaryInfC{$J(P,Q):C(M,N,P)$}
\end{bprooftree}
\end{center}
\bigskip

As we can see, the traditional approach is based on the assumption that there is only one basic proof, the reflexive one. Every proof is then based upon the reflexivity. The complexity of the equality expression does not matter. If two terms $a,b : A$ are really propositionally equal, the constructor $J$ will construct a term $p : Id_{A}(a,b)$. It is incredible to think that a reflexive base is capable of building complexes proofs. Nevertheless, the beauty of the identity type comes with a setback: sometimes the constructor $J$ is not easy to use. The problem is the high amount of information. To use $J$ to build a type, one needs to come up with a suitable reason that justifies the equality, i.e., suitable $x, y : A$ and $z : Id_{A}(x,y)$ that build the correct $C(x,y,z)$. After obtaining the correct $C$, one sometimes realize that $C(x,x,r(x))$ is a type other than $Id_{A}(x,x)$. If $C(x,x,r(x)) \equiv Id_{A}(x,x)$, $Q$ would be the trivial reflexive term, $r(x)$, and the construction would be complete. But if it is not the case, one will need to use the constructor $J$ recursively to build the term $Q$. Of course it means that one will need to find yet another reason to build a suitable $C'(x,y,z)$. Since finding reasons can be cumbersome, proofs using $J$ can be really difficult sometimes. It can get even worse: one may need lots of steps and applications of $J$ before obtaning $C(x,x,r(x)) \equiv Id_{A}(x,x)$. On the other hand, after completing the proof, the correctness can be easily checked by computers. The real problem is how hard it is to find the correct reason and the number of recursive steps.

Simplifying this process is one of the motivations of the path-based approach. The path-based identity type will be constructed the same way that the pathless one had been, i.e., using natural deductions to formalize each rule. We start with the formation and introduction:X

\bigskip
\begin{center}
\begin{bprooftree}
\AxiomC{$A$ type}
\AxiomC{$a : A$}
\AxiomC{$b : A$}
\RightLabel{$Id - F$}
\TrinaryInfC{$Id_{A}(a,b)$ type}
\end{bprooftree}
\begin{bprooftree}
\AxiomC{$a =_{s} b : A$}
\RightLabel{$Id - I_{1}$}
\UnaryInfC{$s(a,b) : Id_{A}(a,b)$}
\end{bprooftree}
\end{center}
\bigskip

No surprises about the $Id - F$, since it is equal to the pathless approach. The $Id-I_{1}$ is where lies the first difference: if we have a computational path between $a,b : A$, then the path is a evidence of the equality of these terms (since from $a$ we can apply a sequence of rewrites, represented by $s$, to reach $b$). Therefore, we introduce the term $s(a,b) : Id_{A}(a,b)$. As an example of how such a proof term is built up, let us recall a pair of terms used in a previous example:

\begin{example}\label{examplepath}
The term $M\equiv(\lambda x.(\lambda y.yx)(\lambda w.zw))v$ is $\beta\eta$-equal to $N\equiv zv$ because of the sequence:\\
$(\lambda x.(\lambda y.yx)(\lambda w.zw))v, \quad  (\lambda x.(\lambda y.yx)z)v, \quad   (\lambda y.yv)z , \quad zv$\\
Now, taking this sequence into a path leads us to the following:\\
The first is equal to the second based on the grounds:\\
$\eta((\lambda x.(\lambda y.yx)(\lambda w.zw))v,(\lambda x.(\lambda y.yx)z)v)$\\
The second is equal to the third based on the grounds:\\
$\beta((\lambda x.(\lambda y.yx)z)v,(\lambda y.yv)z)$\\
Now, the first is equal to the third based on the grounds:\\
$\tau(\eta((\lambda x.(\lambda y.yx)(\lambda w.zw))v,(\lambda x.(\lambda y.yx)z)v),\beta((\lambda x.(\lambda y.yx)z)v,(\lambda y.yv)z))$\\
Now, the third is equal to the fourth one based on the grounds:\\
$\beta((\lambda y.yv)z,zv)$\\
Thus, the first one is equal to the fourth one based on the grounds:\\
$\tau(\tau(\eta((\lambda x.(\lambda y.yx)(\lambda w.zw))v,(\lambda x.(\lambda y.yx)z)v),\beta((\lambda x.(\lambda y.yx)z)v,(\lambda y.yv)z)),\beta((\lambda y.yv)z,zv)))$.
\end{example}

There is another introduction rule:

\bigskip
\begin{center}
\begin{bprooftree}
\AxiomC{$a =_{s} b : A$ }
\AxiomC{$a =_{t} b : A$}
\AxiomC{$s = _{z} t : Id_{A}(a,b)$}
\RightLabel{$Id - I_{2}$}
\TrinaryInfC{$s(a,b) =_{\xi (z)} t(a,b) : Id_{A}(a,b)$}
\end{bprooftree}
\end{center}

\bigskip

This introduction brings about an interesting aspect, since it introduces an important construction: definitional equality between paths, leading to the possibility of constructing paths between paths. Since a computational path is also a computational object, it is natural to think that we can estabilish the equality of two paths. Since in our approach the evidence of equality is given by paths, then the equality of two paths must be estabilished by another path. If the paths are equal, then this rule indicates that the identity terms will be also equal. The existence of paths between paths is essential, since it turns possible the existence of higher structures (as we will see further in this work). The first elimination follows:

\bigskip
\begin{center}
\begin{bprooftree}
\alwaysNoLine
\AxiomC{$m : Id_{A}(a,b)$ }
\AxiomC{$[a =_{g} b : A]$}
\UnaryInfC{$h(g) : C$}
\alwaysSingleLine
\RightLabel{$Id - E_{1}$}
\BinaryInfC{$REWR(m, \acute{g}.h(g)) : C$}
\end{bprooftree}
\end{center}

\bigskip

The elimination works as follows: if from the equality of $a$ and $b$ (estabilished by a path $g$) one can build a term $h(g) : C$, then from another proof of the equality of these two terms (estabilished by $m$) one also should be capable of building a term of type $C$. That idea is captured by the constructor $REWR$. It receives $m$ and the $\acute{g}.h(g)$, where $\acute{g}$ is an abstraction. More specifically, $\acute{g}$ is an abstraction over the variable $g$, for which the main rules of conversion of $\lambda$-abstraction hold. To simulate the fact that from $m$ one should be capable of building a term of $C$, one should be capable of applying $m$ in the expression $\acute{g}.h(g)$, obtaining $h(m/g)$ (i.e., the expression $h$ with the free occurrences of $g$ replaced with the term $m$, so that $h(m) : C$. Of course, this computation should be formalized by a rule. That is exactly what we will do, but before that, let us s show the second elimination:

\bigskip
\begin{center}
\begin{bprooftree}
\alwaysNoLine
\AxiomC{$p =_{r} q : Id_{A}(a,b)$ }
\AxiomC{$[a =_{g} b : A]$}
\UnaryInfC{$h(g) : C$}
\alwaysSingleLine
\RightLabel{$Id - E_{2}$}
\BinaryInfC{$REWR(p, \acute{g}.h(g)) =_{\mu (r)} REWR(q, \acute{g}.h(g)) : C$}
\end{bprooftree}
\end{center}

\bigskip

This rule is similar to what happened in the second introduction rule. If we have two equal paths (equality estabilished by another path), then if we apply the elimination in each of these paths we should obtain equal $REWR$ terms. Now, we define the computation rule that we have just mentioned: 

\bigskip
\begin{center}
\begin{bprooftree}
\AxiomC{$a =_{m} b : A$}
\RightLabel{$Id - I_{1}$}
\UnaryInfC{$m(a,b) : Id_{A}(a,b)$}
\alwaysNoLine
\AxiomC{$[a =_{g} b : A]$}
\UnaryInfC{$h(g) : C$}
\alwaysSingleLine
\RightLabel{$Id - E_{1}$ \quad $\rhd_\beta$}
\BinaryInfC{$REWR(m, \acute{g}.h(g)) : C$}
\end{bprooftree}
\begin{bprooftree}
\AxiomC{$a =_{m} b : A$}
\alwaysNoLine
\UnaryInfC{$h(m/g):C$}
\end{bprooftree}
\end{center}
\bigskip

This rules formalizes what we have described. It creates a reduction rule for the term $REWR(m, \acute{g}.h(g))$. This reduction is just the application of $m$ in $\acute{g}.h(g)$, obtaining $h(m/g)$. Since this reduction is similar to the $\beta$ one of the $\lambda\beta\eta$-equality, we chose to call it $\beta$-reduction for computational paths. There is another reduction rule:

\bigskip

\begin{center}
\begin{bprooftree}
\AxiomC{$e : Id_{A}(a,b)$}
\AxiomC{$[a =_{t} b : A]$}
\RightLabel{$Id - I_{1}$}
\UnaryInfC{$t(a, b) : Id_{A}(a, b)$}
\RightLabel{$Id - E_{1}$ \quad  $\rhd_{\eta}$ \quad $e : Id_{A}(a,b)$}
\BinaryInfC{$REWR(e, \acute{t}.t(a,b)) : Id_{A}(a,b)$}
\end{bprooftree}
\end{center}

\bigskip

This rule is proposed to handle the trivial case, i.e., when the term originated by $a =_{t} b$ is the simplest one: the term $t(a,b) : Id_{A}(a,b)$. Since the term $e$ is already a term of type $Id_{A}(a,b)$, we just reduce $REWR(e, \acute{t}.t(a,b)) : Id_{A}(a,b)$ directly to $e$. We chose to call it $\eta$-reduction because the $\eta$-reduction of the $\lambda\beta\eta$-equality also handles a trivial case, reducing $\lambda x.Mx$ to just $M$. This rule is the last one of our approach. 

\subsection{Path-based constructions}

The objective of this subsection is to show how to use in practice the rules that we have just defined. The idea is to show construction of terms of some important types. The constructions that we have chosen to build are the reflexive, transitive and symmetric type of the identity type. Those were not random choices. The main reason is the fact that reflexive, transitive and symmetric types are essential to the process of building a groupoid model for the identity type \cite{hofmann1}. As we shall see, these constructions come naturally from simple computational paths constructed by the application of axioms of the equality of type theory. In contrast, we will also show that constructing the transitivity using the operator $J$ can be a little complicated. With that, we hope to make the simplicity of our approach clear.

Before we start the constructions, we think that it is essential to understand how to use the eliminations rules. The process of building a term of some type is a matter of finding the right reason. In the case of $J$, the reason is the correct $x,y : A$ and $z : Id_{A}(a,b)$ that generates the adequate $C(x,y,z)$. In our approach, the reason is the correct path $a =_{g} b$ that generates the adequate $h(g) : Id(a,b)$. 

One could find strange the fact that we need to prove the reflexivity. Nevertheless, just remember that our approach is not based on the idea that reflexivity is the base of the identity type. As usual in type theory, a proof of something comes down to a construction of a term of a type. In this case, we need to construct a term of type $\Pi_{(a : A)}Id_{A}(a,a)$. The reason is extremely simple: from a term $a : A$, we obtain the computational path $a =_{\rho} a : A$:

\bigskip

\begin{center}
\begin{bprooftree}
\AxiomC{$[a : A]$}
\UnaryInfC{$a =_{\rho} a : A$}
\RightLabel{$Id - I_{1}$}
\UnaryInfC{$\rho(a,a) : Id_{A}(a,a)$}
\RightLabel{$\Pi-I$}
\UnaryInfC{$\lambda a.\rho(a,a) : \Pi_{(a : A)}Id_{A}(a,a)$}
\end{bprooftree}
\end{center}

\bigskip

As one can see, we did not need to apply the elimination rule. The reason for that is the fact that the reflexive path has already given us a term of the desired type.

The second proposed construction is the symmetry. We need to construct a term of type  $\Pi_{(a : A)}\Pi_{(b : A)}(Id_{A}(a,b) \rightarrow Id_{A}(b,a))$. As expected, we need to find a suitable reason. Starting from $a =_{t} b$, we could look at the axioms of \emph{definition 2.3} to plan our next step. One of those axioms makes the symmetry clear: the $\sigma$ axiom. If we apply $\sigma$, we will obtain $b =_{\sigma(t)} a$. Now, it is just a matter of applying the elimination: 

\bigskip

\begin{center}
\begin{bprooftree}
\AxiomC{$[p : Id_{A}(a,b)]$}
\AxiomC{[$a =_{t} b : A$]}
\UnaryInfC{$b =_{\sigma(t)} a : A$}
\RightLabel{$Id - I_{1}$}
\UnaryInfC{$(\sigma(t))(b,a) : Id_{A}(b,a)$}
\RightLabel{$Id - E_{1}$}
\BinaryInfC{$REWR(p,\acute{t}.(\sigma(t))(b,a)) : Id_{A}(b,a)$}
\UnaryInfC{$\lambda p.REWR(p, \acute{t}.(\sigma(t))(b,a)) : Id_{A} (a,b) \rightarrow Id_{A}(b,a)$}
\RightLabel{$\Pi-I$}
\UnaryInfC{$\lambda b. \lambda p.REWR(p,\acute{t}.(\sigma(t))(b,a)) :  \Pi_{(b : A)}(Id_{A} (a,b) \rightarrow Id_{A}(b,a))$}
\RightLabel{$\Pi-I$}
\UnaryInfC{$\lambda a.\lambda b. \lambda p.REWR(p, \acute{t}.(\sigma(t))(b,a)) :  \Pi_{(a : A)}\Pi_{(b : A)}(Id_{A} (a,b) \rightarrow Id_{A}(b,a))$}
\end{bprooftree}
\end{center}

\bigskip

The third and last construction will be the transitivity. The transitivity will be a special case, since we will also show the construction using the pathless approach, i.e., using the operator $J$. The objective is to show the difference of complexity in the process of finding a suitable reason. Both approaches have the objective of constructing a term for the type $\Pi_{(a : A)}\Pi_{(b : A)}\Pi_{(c : A)} (Id_{A}(a,b) \rightarrow Id_{A}(b,c) \rightarrow Id_{A}(a,c))$.

Let's start with the construction based on $J$. The proof will be based on the one found in \cite{hott}. The difference is that instead of defining induction principles for $J$ based on the elimination rules, we will use the rule directly. The complexity is the same, since the proofs are two forms of presenting the same thing and they share the same reasons. As one should expect, the first and main step is to find a suitable reason. In other words, we need to find suitable $x, y : A$ and $z : Id_{A}(a,b)$ to construct an adequate $C(x,y,z)$. This first step is already problematic. Different from our approach, where one starts from a path and applies intuitive equality axioms to find a suitable reason, there is no clear point of how one should prooceed to find a suitable reason for the construction  based on $J$. In this case, one should rely on intuition and make attempts until one finds out the correct reason. As one can check in \cite{hott}, a suitable reason would be $x : A, y : A, - : Id_{A}(x,y)$ and $C(x,y,z) \equiv Id_{A}(y,c) \rightarrow Id_{A}(x,c)$. The symbol $-$ indicates that $z$ can be anything, i.e., the choice of $z$ will not matter. Looking closely, the proof is not over yet. The problem is the type of $C(x,x,r(x))$. With this reason, we have that $C(x,x,r(x)) \equiv Id_{A}(x,c) \rightarrow Id_{A}(x,c)$. Therefore, we cannot assume that $Q(x) : Id_{A}(x,c) \rightarrow Id_{A}(x,c)$ is the term $r(x)$. The only way to proceed is to apply again the constructor $J$ to build the term $Q(x)$. It means, of course, that we will need to find yet another reason to build this type. This second reason is given by  $x : A, y : A, - : Id_{A}(x,y)$ and $C'(x,y,z) \equiv Id_{A}(x,y)$. In that case, $C'(x,x,r(x)) = Id_{A}(x,x)$. We will not need to use $J$ again, since now we have that $r(x) : Id_{A}(x,x)$. Then, we can construct $Q(x)$:

\bigskip
\begin{center}
\begin{bprooftree}
\alwaysNoLine
\AxiomC{$x:A$}
\AxiomC{$c:A$}
\AxiomC{$q:Id_{A}(x,c)$}
\AxiomC{$[x:A]$}
\UnaryInfC{$r(x):Id_{A}(x,x)$}
\AxiomC{$[x:A,y:A,- : Id_{A}(x,y)]$}
\UnaryInfC{$Id_{A}(x,y)$ type} 
\RightLabel{$Id - E$ \quad}
\alwaysSingleLine
\QuinaryInfC{$J(q,r(x)):Id_{A}(x,c)$}
\UnaryInfC{$\lambda q.J(q,r(x)): Id_{A}(x,c) \rightarrow Id_{A}(x,c)$}
\end{bprooftree}
\end{center}
\bigskip

Since we have constructed a term to act as the $Q(x)$ of the elimination rule, we can finally obtain the desired term:

\bigskip
\begin{tiny}
\begin{center}
\begin{bprooftree}
\alwaysNoLine
\AxiomC{$a:A$}
\AxiomC{$b:A$}
\AxiomC{$p:Id_{A}(a,b)$}
\AxiomC{$[x:A]$}
\UnaryInfC{$\lambda q.J(q,r(x)): Id_{A}(x,c) \rightarrow Id_{A}(x,c)$}
\AxiomC{$[x:A,y:A,- : Id_{A}(x,y)]$}
\UnaryInfC{$Id_{A}(y,c) \rightarrow Id_{A}(x,c)$ type} 
\RightLabel{$Id - E$ \quad}
\alwaysSingleLine
\QuinaryInfC{$J(p,\lambda q.J(q,r(x))):  Id_{A}(b,c) \rightarrow Id_{A}(a,c)$}
\UnaryInfC{$\lambda p.J(p,\lambda q.J(q,r(x))): Id_{A}(a,b) \rightarrow Id_{A}(b,c) \rightarrow Id_{A}(a,c)$}
\RightLabel{$\Pi-I$}
\UnaryInfC{$\lambda c.\lambda p.J(p,\lambda q.J(q,r(x))):  \Pi_{(c : A)}(Id_{A}(a,b) \rightarrow Id_{A}(b,c) \rightarrow Id_{A}(a,c))$}
\RightLabel{$\Pi-I$}
\UnaryInfC{$\lambda b.\lambda c.\lambda p.J(p,\lambda q.J(q,r(x))):  \Pi_{(b : A)}\Pi_{(c : A)}(Id_{A}(a,b) \rightarrow Id_{A}(b,c) \rightarrow Id_{A}(a,c))$}
\RightLabel{$\Pi-I$}
\UnaryInfC{$\lambda a.\lambda b.\lambda c.\lambda p.J(p,\lambda q.J(q,r(x))):  \Pi_{(a : A)}\Pi_{(b : A)}\Pi_{(c : A)}(Id_{A}(a,b) \rightarrow Id_{A}(b,c) \rightarrow Id_{A}(a,c))$}
\end{bprooftree}
\end{center}
\end{tiny}
\bigskip

This construction is an example that makes clear the difficulties of working with the pathless model. We had to find two different reasons and use two applications of the elimination rule. Another problem is the fact that the reasons were not obtained by a fixed process, like the applications of axioms in some entity of type theory. They were obtained purely by the intuition that a certain $C(x,y,z)$ should be capable of constructing the desired term. For that reason, obtaining these reasons can be troublesome. 

We finish our constructions by giving the path-based construction of the transitivity. The first step, as expected, is to find the reason. Since we are trying to construct the transitivity, it is natural to think that we should start with paths $a =_{t} b$ and $b =_{u} c$ and then, from these paths, we shoud conclude that there is a path $z$ that estabilishes that $a =_{z} c$. To obtain $z$, we could try to apply the axioms of \emph{definition 2.3}. Looking at the axioms, one is exactly what we want: the axiom $\tau$. If we apply $\tau$ to  $a =_{t} b$ and $b =_{u} c$, we will obtain a new path $\tau(t,u)$ such that $a = _{\tau(t,u)} c$. Using that construction as the reason, we obtain the following term:

\bigskip
\begin{tiny}
\begin{center}
\begin{bprooftree}
\AxiomC{$[w(a,b) : Id_{A}(a,b)]$}
\AxiomC{$s(b,c) : Id_{A}(b,c)$}
\AxiomC{$[a =_{t} b]$}
\AxiomC{$[b =_{u} c]$}
\BinaryInfC{$a =_{\tau(t,u)} c$}
\RightLabel{$Id - I_{1}$}
\UnaryInfC{$(\tau (t,u))(a,c) : Id_{A}(a,c)$}
\RightLabel{$Id - E_{1}$}
\BinaryInfC{$REWR(s(b,c),\acute{u}(\tau (t,u))(a,c)) : Id_{A}(a,c)$}
\RightLabel{$Id - E_{1}$}
\BinaryInfC{$REWR(w(a,b),\acute{t}REWR(s(b,c),\acute{u}(\tau (t,u))(a,c))) : Id_{A}(a,c)$}
\UnaryInfC{$\lambda s.REWR(w(a,b),\acute{t}REWR(s(b,c),\acute{u}(\tau (t,u))(a,c))) : Id_{A}(b,c) \rightarrow Id_{A}(a,c)$}
\UnaryInfC{$\lambda w.\lambda s.REWR(w(a,b),\acute{t}REWR(s(b,c),\acute{u}(\tau (t,u))(a,c))) : Id_{A}(a,b) \rightarrow Id_{A}(b,c) \rightarrow Id_{A}(a,c)$}
\RightLabel{$\Pi-I$}
\UnaryInfC{$\lambda c.\lambda w.\lambda s.REWR(w(a,b),\acute{t}REWR(s(b,c),\acute{u}(\tau (t,u))(a,c))) :  \Pi_{(c : A)}(Id_{A}(a,b) \rightarrow Id_{A}(b,c) \rightarrow Id_{A}(a,c))$}
\RightLabel{$\Pi-I$}
\UnaryInfC{$\lambda b. \lambda c.\lambda w.\lambda s.REWR(w(a,b),\acute{t}REWR(s(b,c),\acute{u}(\tau (t,u))(a,c))) :  \Pi_{(b : A)}\Pi_{(c : A)}(Id_{A}(a,b) \rightarrow Id_{A}(b,c) \rightarrow Id_{A}(a,c))$}
\RightLabel{$\Pi-I$}
\UnaryInfC{$\lambda a. \lambda b. \lambda c.\lambda w.\lambda s.REWR(w(a,b),\acute{t}REWR(s(b,c),\acute{u}(\tau (t,u))(a,c))) :   \Pi_{(a : A)}\Pi_{(b : A)}\Pi_{(c : A)}(Id_{A}(a,b) \rightarrow Id_{A}(b,c) \rightarrow Id_{A}(a,c))$}
\end{bprooftree}
\end{center}
\end{tiny}
\bigskip

As one can see, each step is just straightfoward applications of introduction, elimination rules and abstractions. The only idea behind this construction is just the simple fact that the axiom $\tau$ guarantees the transitivity of paths. If one compare the reason of this construction to the one that used $J$, one can clearly conclude that the reason of the path-based approach was obtained more naturally.

\section{The Groupoid Model}

One of the milestones of the study of the usual identity type was the discovery that the identity type induces an algebraic structure. The structure that it induces is known as groupoid and this connection was originally proposed by Hofmann \& Streicher (1994) \cite{hofmann1}. As proposed in \cite{hofmann1}, the identity type obeys the groupoid laws in a special sense: the equations hold up only to propositional equality. Since all the terms that forms the groupoid were constructed using the operator $J$, the proof that the usual identity type induces a groupoid is not valid for our path-based approach. With that in mind, the objective of this section is to show that our path-based identity type also induces a groupoid structure. Furthermore, our groupoid laws will also hold up only to propositional equality. 

Before we show the groupoid laws, understanding the existence of a path between paths will be crucial. Since computational paths are terms of a type (terms of the identity type specifically), all the axioms of \emph{definition 2.3} apply to any path. Then, what will be a path between paths? This is better explained with a simple example. Consider a generic path $a =_{s} b : A$  and its term $s(a,b) : Id_{A}(a,b)$. Looking at the axioms of \emph{definition 2.3}, one can apply the axiom $\rho$ in $s$, obtaining $s =_{\rho} s$. This path will be the term $\rho (s,s) : Id_{Id_A(a,b)}(s,s)$. That way, $\rho(s,s)$ is a simple example of paths between paths. We are interested in more important examples. Sometimes a path has some redundancies and this redundancies can be reduced, resulting in a new path. Since this new path is just the old one without redundancies, they should be considered propositionally equal, i.e., there should be a path connection these two. Again, it is explained better with some examples:

\begin{example}
Consider a path $a =_{t} b : A$. It is possible to apply the axiom $\sigma$, obtaining $b =_{\sigma(t)} a$ in the process. If we apply again $\sigma$, we will obtain $a = _{\sigma(\sigma(t))} b$. Since we have just inverted the path two times in a row, the path $\sigma(\sigma(t))$ is just a redundant form of the path $t$. Then, $\sigma(\sigma(t))$ should be reduced (or rewrited) to $t$, i.e., there should exist a path estabilishing that $\sigma(\sigma(t))$ and $t$ are propositionally equal. 
\end{example}

\begin{example}
Consider the reflexive path $a =_{\rho} a : A$. We can apply $\sigma$, obtaining $a =_{\sigma(\rho)} a$. Since we applied $\sigma$ in a trivial reflexive path, $\sigma(\rho)$ is just a redundant form of $\rho$. Then, $\sigma(\rho)$ should be reduced to $\rho$ and there should be a path between these two paths.
\end{example}

\begin{example}
Consider the path $a =_{t} b : A$. If we apply $\sigma$, we will obtain $b =_{\sigma(t)} a$. We can take now both $t$ and $\sigma(t)$ and apply the axiom $\tau$, obtaining $a =_{\tau(t,\sigma(t))} a$. Since we have just applied the transitivity to a path and its inverse, the result can be reduced to the reflexivity $\rho$.
\end{example}

The examples show three distinct cases which redundant paths can be reduced to a path without redundancies. We have also concluded that, in each case, there sould be a path estabilishing the equality of the redundant path and the equivalent one free of redundancies. These examples were simple ones involving only the axioms $\tau$, $\sigma$ and $\rho$. Since the equality theory of type theory has a total of seven axioms, as one can check in \emph{definition 2.3}, the combination of these axioms should cause the appearance of many different redundancies. Since we have just showed the simple ones, mapping all these redundancies would be a hard task. Fortunately, this mapping has already been made by De Oliveira (1995) \cite{Anjo1}.

The analysis of the redundancies generated by the axioms of equality was the main proposal of \cite{Anjo1}. In that work, De Oliveira identifies all the redundancies cases and creates rules to reduce these redundancies to a term free of redundancies. These set of all reduction rules forms a system called $LND_{EQ}-TRS$. As we have just seen in the examples, each reduction rules estabilishes that one path (the redundant one) can be rewrited into another one without redundancies, i.e., there is a path between these two paths. The full set of rules of $LND_{EQ}-TRS$ has a total of 39 reduction rules \cite{Anjo1}. (In \cite{Ruy1}, identifiers for each of those definitional equalities were added, so that while the 1st level identifiers for definitional equalities $\beta$, $\eta$, $\xi$, etc., are the basis for computational paths of this level, identifiers such as $tt$, $ss$, $sr$, etc., will serve the same role for the next level up.) Fortunately, we are only interested in a subset of these rules, namely the ones that involve the axioms of transitivity, reflexivity and symmetry. Each rule has a name and for the sake of simplicity, the path that the rule induces will receive the same name. Here follow all the rules that interest us and their deductions \cite{Anjo1}:

\begin{itemize}
\item Rules involving $\sigma$ and $\rho$

\bigskip

\begin{prooftree}
\AxiomC{$x =_{\rho} x : A$}
\RightLabel{\quad $\rhd_{sr}$ \quad $x =_{\rho} x : A$}
\UnaryInfC{$x =_{\sigma(\rho)} x : A$}
\end{prooftree}

\begin{prooftree}
\AxiomC{$x =_{r} y : A$}
\UnaryInfC{$y =_{\sigma(r)} x : A$}
\RightLabel{\quad $\rhd_{ss}$ \quad $x =_{r} y : A$}
\UnaryInfC{$x =_{\sigma(\sigma(r))} y : A$}
\end{prooftree}

\bigskip

From these reductions, we obtain the following paths:

$\sigma(\rho) =_{sr} \rho$

$\sigma(\sigma(r)) =_{ss} r$
\bigskip

\item Rules involving $\tau$

\bigskip
\begin{prooftree}
\AxiomC{$x =_{r} y : A$}
\AxiomC{$y =_{\sigma(r)} x : A$}
\RightLabel{\quad  $\rhd_{tr}$ \quad $x =_{\rho} x : A$}
\BinaryInfC{$x =_{\tau(r,\sigma(r))} x : A$}
\end{prooftree}

\begin{prooftree}
\AxiomC{$y =_{\sigma(r)} x : A$}
\AxiomC{$x =_{r} y : A$}
\RightLabel{\quad $\rhd_{tsr}$ \quad $y =_{\rho} y : A$}
\BinaryInfC{$y =_{\tau(\sigma(r),r)} y : A$}
\end{prooftree}

\begin{prooftree}
\AxiomC{$x =_{r} y : A$}
\AxiomC{$y =_{\rho} y : A$}
\RightLabel{\quad $\rhd_{trr}$ \quad $x =_{r} y : A$}
\BinaryInfC{$x =_{\tau(r,\rho)} y : A$}
\end{prooftree}

\begin{prooftree}
\AxiomC{$x =_{\rho} x : A$}
\AxiomC{$x =_{r} y : A$}
\RightLabel{\quad $\rhd_{tlr}$ \quad $x =_{r} y : A$}
\BinaryInfC{$x =_{\tau(\rho,r)} y : A$}
\end{prooftree}

\bigskip

Obtained paths:

$\tau(r,\sigma(r)) =_{tr} \rho$

$\tau(\sigma(r),r) =_{tsr} \rho$

$\tau(r,\rho) =_{trr} r$

$\tau(\rho,r) =_{tlr} r$

\bigskip

\item Rule involving $\tau$ and $\tau$

\begin{prooftree}
\hskip - 155pt
\AxiomC{$x =_{t} y : A$}
\AxiomC{$y =_{r} w : A$}
\BinaryInfC{$x =_{\tau(t,r)} w : A$}
\AxiomC{$w =_{s} z : A$}
\BinaryInfC{$x =_{\tau(\tau(t,r),s)} z : A$}
\end{prooftree}

\begin{prooftree}
\hskip 4cm
\AxiomC{$x =_{t} y : A$}
\AxiomC{$y=_{r} w : A$}
\AxiomC{$w=_{s} z : A$}
\BinaryInfC{$y =_{\tau(r,s)} z : A$}
\LeftLabel{$\rhd_{tt}$}
\BinaryInfC{$x =_{\tau(t,\tau(r,s))} z : A$}
\end{prooftree}

\bigskip

Obtained path:

$\tau(\tau(t,r),s) =_{tt} \tau(t, \tau(r,s))$
\end{itemize}

These paths are all what we need to show that our path-based identity type induces a groupoid structure. Before we show the groupoid laws and the construction of the terms, we'd like to add that the $LND_{EQ}-TRS$ system of rules is terminating and confluent. Terminating in the sense that from a generic path $s$ one can apply the rules until one obtain a path free of redundancies. Confluente in the sense that if a path $s$ can be reduced to $m$ and the same $s$ can also be reduced to $n$, then there is a $r$ such that $m$ and $n$ can be reduced to $r$. The proof of these two properties can be found in \cite{Anjo1,Ruy2,Ruy3,RuyAnjolinaLivro}.

\subsection{The Groupoid Laws}

The groupod model of a type is the idea that the identity type induces a groupoid structure on every type $A$. Originally proposed by \cite{hofmann1}, the model is based on the fact that $J$ is capable of constructing terms for the following types \cite{hofmann1}:

\begin{itemize}

\item $Id_{Id_{A}(x,z)}(trans(trans(p,q), r), trans(p, trans(q,r)))$
\item $Id_{Id_{A}(x,y)}(trans(refl(x), r), r)$
\item $Id_{Id_{A}(x,y)}(trans(r,refl(x)), r)$
\item $Id_{Id_{A}(y,y)}(trans(symm(r),r), refl(x))$
\item $Id_{Id_{A}(x,x)}(trans(r, symm(r)), refl(x))$
\item $Id_{Id_{A}(x,y)}(symm(symm(r)),r)$
\end{itemize}

One important detail is, as one can see, that these equalities do not hold ``on the nose", so to speak. They hold only up to propositional equality, since every term is an inhabitant of an identity type. 

The terms $trans$, $symm$ and $refl$, as one can expect, represent the transitivity, symmetry and reflexivity. (To be precise, only $refl$ is part of the language of terms of identity types in the traditional approach, which leaves open the question as to where do $symm$ and $trans$ come from.) In our path-based interpretation, we write $trans$ as $\tau$, $symm$ as $\sigma$ and $refl$ as $\rho$, and they are all identifiers for definitional equalities. Since we already have obtained all the paths we need through the reduction rules, our path-based identity type can construct these terms in a rather natural fashion. For example, consider the path $tt$ between paths, i.e.\ $\tau(\tau(p,q),r) =_{tt} \tau(p, \tau(q,r))$, then we have:

\bigskip
\begin{center}
\begin{bprooftree}
\AxiomC{ $\tau(\tau(p,q),r) =_{tt} \tau(p, \tau(q,r)) : Id_{A}(x,z)$}
\RightLabel{$Id - I_{1}$}
\UnaryInfC{$tt((\tau(\tau(p,q),r), \tau(p, \tau(q,r))) : Id_{Id_{A}(x,z)}(((\tau(\tau(p,q),r), \tau(p, \tau(q,r)))$}
\end{bprooftree}
\end{center}
\bigskip

In an analogous way, we can obtain the remaining terms. Hence, we will have all the terms we need:

\begin{itemize}

\item $tt((\tau(\tau(p,q),r), \tau(p, \tau(q,r))) : Id_{Id_{A}(x,z)}(((\tau(\tau(p,q),r), \tau(p, \tau(q,r)))$
\item $tlr (\tau(\rho,r), r) : Id_{Id_{A}(x,y)}(\tau(\rho,r), r)$
\item $trr (\tau(r,\rho), r) : Id_{Id_{A}(x,y)}(\tau(r,\rho), r)$
\item $tsr (\tau(\sigma(r),r), \rho) : Id_{Id_{A}(y,y)}(\tau(\sigma(r),r), \rho)$
\item $tr (\tau(r,\sigma(r)), \rho) : Id_{Id_{A}(x,x)}(\tau(r,\sigma(r)), \rho)$
\item $ss (\sigma(\sigma(r)),r) : Id_{Id_{A}(x,y)}(\sigma(\sigma(r)),r)$
\end{itemize}

Since paths only estabilish propositional equality, this will be the same case of the pathless approach: all equalities will hold up to propositional equality. We conclude that, for any type $A$, our path-based identity type induces a groupoid model.

\subsection{Sketch of a Higher Structure}

In this subsection, our objective is to imply that computational paths can induce a higher structure. We will not expose deeper details, since most of what will be presented in this subsection is still work in progress. Nevertheless, we think that what will be exposed here is interesting enough to be part of this work. 

The groupoid model that we have just exposed was based on the fundamental idea that a computational path, as term of a type, is also a computational object. Thefore, it is reasonable to think of paths between paths. Based on that, we can extend the idea and think about paths between paths of paths. We can take it even more further, thinking about paths between paths of paths of paths. In fact, we can be more generic and define the level of a path:

\begin{definition}

The \emph{level} of a path has the following inductive definition:

\begin{itemize}

\item If $a,b : A$ are terms such that $a$ and $b$ are not computational paths, then we say that a path $a =_{s} b$ has level 0.

\item If $r,s: A$ are paths of level $n$, then a path $r =_{\theta} s$ has level $n + 1$.

\end{itemize}

\end{definition}

The idea is that if we have two $n$-level paths $r$ and $s$ and if  $r$ can be rewrited into $s$ by a sequence of rewrites (a rewrite in the same sense that we defined before), then this sequence of rewrites is a $(n+1)$-level path that estabilishes that $r$ and $s$ are propositionally equal. Consider now the following structure \cite{Tom}:

\begin{definition}
Let $n \in \Nat$. An $n$-globular set $X$ is the following diagram of sets and functions: 

\bigskip

\begin{center}
\begin{tikzpicture}[node distance=2.8cm, auto]

\node (P) {$X(n)$};
\node(Q)[right of=P] {$X(n-1)$};
\node (R)[right of =Q] {$\quad ... \quad$};
\node(S)[right of =R] {$X(0)$};

\draw[transform canvas={yshift=0.5ex},->] (P) --(Q) node[above,midway] {$s$};
\draw[transform canvas={yshift=-0.5ex},->](P) -- (Q) node[below,midway] {$t$}; 
\draw[transform canvas={yshift=0.5ex},->] (Q) --(R) node[above,midway] {$s$};
\draw[transform canvas={yshift=-0.5ex},->](Q) -- (R) node[below,midway] {$t$}; 
\draw[transform canvas={yshift=0.5ex},->] (R) --(S) node[above,midway] {$s$};
\draw[transform canvas={yshift=-0.5ex},->](R) -- (S) node[below,midway] {$t$}; 

\end{tikzpicture}
\end{center}

\bigskip

The diagram has to obey the following equations for all $x \in \{2, . . . ., n\}$ and $x\in X(m)$:

\begin{center}
$s(s(x)) = s(t(x))$, \quad $t(s(x)) = t(t(x))$
\end{center}

\end{definition}

Computational paths can form a globular-set structure. To see this, consider $X(n)$ as the set of $n$-level paths between two ($n-1$)-level paths (for $n = 0$, just consider two non-path objects) and for any path $x =_{m} y$, consider that $s(m) = x$ and $t(m) = y$. That way, if $n \geq 2$ and $m \in X(n)$, then $s(m) \in X(n - 1)$ and $t(m) \in X(n-1)$. Now, if we consider two $n$-level paths $t =_{\theta} m$, and $t = _{\alpha} m$ and a $(n+1)$-level path $\theta =_{\phi} \alpha$, then $s(s(\phi)) = s(\theta) = t = s(\alpha) = s(t(\phi))$ and $t(s(\phi)) = t(\theta) = m = t(\alpha) = t(s(\phi))$. All globular-set conditions hold. Then, we can think that computational paths form a globular set structure all up to infinity, i.e., a $\infty$-globular-set.

Now that we already know that paths form a globular-set structure, we would like to study the redundancies caused by a $n$-level path. We have exposed the existence of a system known as $LND_{EQ}-TRS$, which estabilishes all rules that solves redundancies between $0$-level paths. Since there are redundancies between $0$-level paths, it is reasonable to think that there will also exist redundancies between $n$-level ones. In a analogous way, we could define a $LND_{EQ}-TRS_{n}$ system which have all the rules that resolve redundancies between $n$-level paths. Neverless, the complete understanding of $n$-level redundancies and what possible rules $LND_{EQ}-TRS_{n}$ should have is still work in progress. Nonetheless, we hope to achieve some important results involving higher structures. Our hopes are based on the following analogy: if the study of $0$-level paths and the  $LND_{EQ}-TRS$ resulted in the fact that our path approach induces a groupoid, result that is on par with known results for the traditional identity type, then we strongly believe that our multilevel approach will induce a higher structure similar to the one induced by the pathless identity type. We are refering to the results obtained in \cite{lumsdaine1,Benno}, which concluded that the traditional identity type induces a higher structure known as weak $\omega$-groupoid. Given the high complexity of such structure, obtaining it using our approach will possibly be the main focus of a future work.

\section{Conclusion}

Inspired by a recent discovery that the propositional equality between terms can be interpreted as a homotopical path, we have revisited the formulation of the intensional identity type, proposing a new approach based on a entity known as {\em computational path}. We have proposed that a computational path $a =_{s} b : A$ is a term $s(a,b)$ of the identity type, i.e., $s(a,b) : Id_{A}(a,b)$, and is formed by a composition of basic rewrites, each with their identifiers taken as constants. We have also developed our approach, showing how the path-based identity type can be constructed and used. In particular, we have showed the simplificity of our elimination rule, showing that it is based on path constructions, which are built from applications of simple axioms of the equality for type theory. To make our point even clearer, we have exposed three path-based constructions. More specifically, constructions that prove the transitivity, reflexivity and symmetry of the identity type. We have also argued that, in our approach, the process of finding the reason that builds the desired term is usually simple and straightfoward. At the same time, in the traditional (or pathless) approach, this is not entirely true, since finding the correct reason can be a cumbersome process.

After estabilishing the foundations of our approach, we analyzed one important structure that the traditional identity type induces: the algebraic structure known as grupoid. Our objective was to show that our approach is on par with the pathless one, i.e., our path-based identity also induces a groupoid sturcture.. To prove that, we have showed that the axioms of equality generate redundancies, which are resolved by paths between paths. We have also exposed that there already exists a system, called $LND_{EQ}-TRS$, that maps those redundancies and proposes rules to reduce them. Using some of these reduction rules, we have obtained all the necessary paths that build the groupoid terms. After obtaining this result, we made a brief sketch of a possible higher structure induced by the path-based identity type. Using paths and deeper level of paths, we have showed that it is possible to form a $\infty$-globular-set. With that, we have laid the groundwork to prove that our path-based identity type induces a weak $\omega$-groupoid structure.

\bibliographystyle{plain}
\bibliography{Biblio}

\end{document}